\documentclass[twocolumn,prl,aps]{revtex4} 
\usepackage{amssymb}
\usepackage{epsfig,amsmath}
\usepackage{graphicx}
\usepackage{dcolumn}
\usepackage{bm}

\newcommand{\beq}{\begin{equation}}
\newcommand{\eeq}{\end{equation}}
\newcommand{\beqa}{\begin{eqnarray}}
\newcommand{\eeqa}{\end{eqnarray}}

\newcommand{\la}{\langle} 
\newcommand{\ra}{\rangle}

\newcommand{\rp}{r_\perp}

\def\lphys#1{{ Laser\ Phys.} {\bf#1}}
\def\nat#1{{ Nature} {\bf#1}}

\def\npho#1{{Nature\ Phot.} {\bf#1}}

\def\OPN#1{{ Opt.\ \& Phot.\ News} {\bf#1}}

\def\oc#1{{ Opt.\ Commun.} {\bf#1}}
\def\ol#1{{ Opt.\ Lett.} {\bf#1}}
\def\opt#1{{ Optica} {\bf#1}}
\def\pla#1{{ Phys.\ Lett. A\/} {\bf#1}}

\def\pra#1{{ Phys.\ Rev. A\/} {\bf#1}}

\def\prd#1{{ Phys.\ Rev. D\/} {\bf#1}}

\def\prl#1{{ Phys.\ Rev.\ Lett.} {\bf#1}}

\begin{document}

\title {Bohr's Complementarity Completed with Entanglement}
\author{X. -F. Qian$^{1,2,3}$}
\email{xiaofeng.qian@rochester.edu}
\author{A. N. Vamivakas$^{1,2,3}$}
\author{J. H. Eberly$^{1,2,3}$}
\affiliation{$^{1}$Center for Coherence and Quantum Optics, \\ 
$^{2}$The Institute of Optics and $^{3}$Department of Physics \& Astronomy\\
University of Rochester, Rochester, New York 14627}
\date{\today }

\begin{abstract}
\end{abstract}

\maketitle
\begin{center}{\large{\bf Introduction} }\end{center}

\noindent{\bf In the Beginning}\quad Ninety years ago in 1927, at an international congress in Como, Italy, Niels Bohr gave an address which is recognized as the first instance in which the term ``complementarity", as a physical concept, was spoken publicly \cite{Bohr-Nature}, revealing Bohr's own thinking about Louis de Broglie's ``duality". Bohr had very slowly accepted duality as a principle of physics: close observation of any quantum object will reveal either wave-like or particle-like behavior, one or the other of two fundamental and complementary features.  Little disagreement exists today about complementarity's importance and broad applicability in quantum science. Book-length scholarly examinations even provide speculations about the relevance of  complementarity in fields as different from physics as biology, psychology and social anthropology, connections which were apparently of interest to Bohr himself (see Jammer \cite{Jammer1}, Murdoch \cite{Murdoch} and Whitaker \cite{Whitaker}). Confusion evident in Como following his talk was not eliminated by Bohr's article \cite{Bohr-Nature}, and complementarity has been subjected to nine decades of repeated examination ever since with no agreed resolution. Semi-popular treatments \cite{SciAm} as well as expert examinations \cite{NatNat, Englert, Menzel, DeZela} show that the topic cannot be avoided, and complementarity retains its central place in the interpretation of quantum mechanics. However, recent approaches by our group \cite{Abouraddy, PhysScr, QMVE, PCT} and others \cite{Plenio, Adesso, JSV, BEG-JAB, ALuis, JB, Khoury} to the underlying notion of coherence now allow us to present a universal formulation of complementarity that may signal the end to the confusion. We demonstrate a new relationship that constrains the behavior of an electromagnetic field (quantum or classical) in the fundamental context of two-slit experiments. We show that entanglement is the ingredient needed to complete Bohr's formulation of complementarity, debated for decades because of its incompleteness. 
\\

\noindent {\bf Frustrating and Frustrated} The audience for the talk in Como included a remarkable array of scientists now known for their significant contributions to the development of quantum theory \cite{names}. Most of them were intrigued to hear what Bohr would have to say, but after the talk there was widespread confusion among them (see \cite{Jammer1}). What exactly had Bohr said?  The long record of frustrating and conflicting opinions about the meaning of complementarity is common knowledge. One asks, exactly what can and should be said about the tension between equally compelling but nevertheless competing particle vs. wave interpretations of phenomena in the natural world? Examples of opposing concepts are readily available, as in the photoelectric effect (light waves behaving as particles) and in electron diffraction (particles behaving as waves). Evidently frustrated himself, Bohr addressed and re-addressed the matter. His writings reveal a number of rephrasings and some would even say reinterpretations of his own ideas. But near the end of his career Bohr reinforced earlier pronouncements (see \cite{Schilpp}), saying simply (for Bohr) that ``... evidence obtained under different conditions cannot be comprehended within a single picture, but must be regarded as {\em complementary} in the sense that only the totality of the phenomena exhausts the possible information about the objects". One then understands, as Whitaker paraphrases Murdoch \cite{Whitaker}, that two factors are critical and must be involved in any correct complementary description - {\em mutual exclusivity} and {\em joint completion}. \\

\begin{figure}[t!]
\includegraphics[width=8cm]{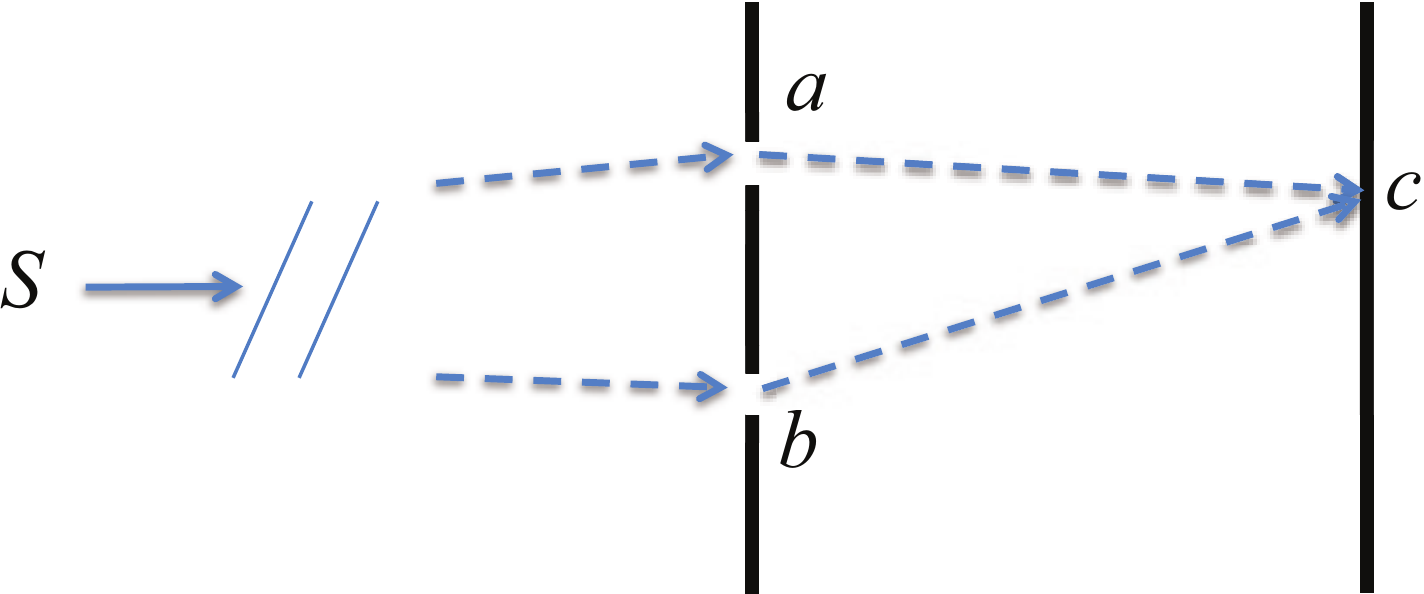}
\caption{The Young two-slit experiment \cite{Young} shows interference fringes on the screen at $c$ where the beams $a$ and $b$ from source $S$ overlap. This is the prototype for all more elaborate scenarios. }
\label{MachZ}
\end{figure}

\noindent {\bf Universally Overlooked.} An important element was for many years not recognized in the earliest discussions. Although the duality between particles and waves, and between rays and waves in optical contexts, was traditionally considered absolute (mutually exclusive), it does not in fact preclude a kind of sharing. An information-theoretic approach by Wootters and Zurek \cite{W and Z} in 1979 brought this fact to the front. In subsequent debate, quantitative formulations began to be found for duality and applied to Bohr's complementarity principle, and almost uncountably many derivations and re-derivations have appeared since 1985 \cite{V2D2proofs}. Almost all of the quantitative approaches have been conceived for experiments using one or another variation of two-channel setups that are historic great-grandchildren of Thomas Young's  \cite{Young} seminal two-slit optical interference experiment (Fig. \ref{MachZ}).
. 

The visibility $V$ of interference fringes (a marker of wave character), and which-path, or which-slit, distinguishability $D$ (a marker of ray or particle character) are both exposed to quantitative observation in the Young scenario. The many related analyses have reported quantified trade-off between visibility $V$ and which-path distinguishability $D$. In fact almost all agree on the validity of the same result, a duality-complementarity inequality:
\beq \label{V2D2}
V^2 + D^2 \le 1.
\eeq

Interested readers who are exasperated and exhausted by so many repeated attempts to understand complementarity more deeply may be asking, if previous treatments have all arrived at (\ref{V2D2}), the same duality formula, what can the present Letter be about? What the present Letter is about is a resolution and completion, as well as generalization, of the discussion, prompted by the slow but gradual recognition that notions perceived to be quantum mechanical in origin, such as entanglement, are in fact equally accessible within the realm of classical wave physics because it also has a vector space foundation. Our group has been at the forefront of these investigations. For some background, one may consult a recent overview of activity in this area \cite{OPN} and four prominent papers \cite{Abouraddy, DeZela, QMVE, Spreeuw}. \\ 

\begin{center}{\large{\bf Theoretical Examination} }\end{center}

\noindent
{\bf Derivation}

\noindent We proceed in the standard double slit context. As sketched in Fig. \ref{MachZ}, a light source illuminates the slits $a$ and $b$ and we consider two transmitted beams that meet on the screen at a point $c$. This analysis will be classical, but the quantum mechanical single photon case obtains the same result.

The optical field that enters the double slit can be written in terms of its spatial, temporal and spin polarization degrees of freedom as:  

\beq \label{incoherent-field}
\vec{E}(\rp,z,t)=A u_a(\rp,z)\vec{\phi}_a(t) + Bu_b(\rp,z)\vec{\phi}_b(t),
\eeq
with $A$ and $B$ being respective field strength amplitudes and containing propagation factors (see below). The spatial dependences of the fields emitted from the two slits carry information of the slits that plays a key role in the interference on the screen. Therefore, we have singled out the spatial components $u_a(\rp,z)$ and $u_b(\rp,z)$, which are unit-normalized orthogonal diffractive functions \cite{B&W}, i.e., $\int  u^*_au_b dxdy=\delta_{ab}$. The remaining degrees of freedom of the field, namely time and spin polarization, are represented by the two vector functions $\vec{\phi}_a$ and $\vec{\phi}_b$, which are chosen to be unit-normalized with $\la\vec{\phi}^*_a\cdot \vec{\phi}_a\ra = \la\vec{\phi}^*_b\cdot \vec{\phi}_b\ra = 1$. Here the dot product symbol $\cdot$ represents inner product of spin vectors and $\la \cdots \ra$ stands for an ensemble average over the stochastic time functions. In general the $\vec{\phi}$ functions are only partially coherent, i.e., $\la\vec{\phi}^*_a\cdot \vec{\phi}_b\ra = \gamma$, where $|\gamma| \le 1$.

In the familiar distant-screen forward-propagation treatment \cite{B&W}, the propagation factors from slits $a$ and $b$ to screen $c$ will only pick up phase differences $\exp(ikr_{ac})$ and $\exp(ikr_{bc})$ (which are absorbed by amplitudes $A$ and $B$). Then the intensity at the screen $c$ is obtained with a standard sinusoidal interference term
\beq
I_c =I_{ac}+I_{bc}+2|\gamma|\sqrt{I_{ac}I_{bc}}\cos[\arg(\gamma A^*B)], \label{I_c}
\eeq
where $I_{ac}=|A|^2$ and $I_{bc}=|B|^2$.

The visibility $V$ of the measured fringes has its usual expression, based on the extreme $\pm 1$ values of the cosine function, which give
\beq \label{V-expression}
V = \frac{I_{max} - I_{min}}{I_{max} + I_{min}} = \frac{2|\gamma|\sqrt{I_{ac}I_{bc}}}{I_{ac}+I_{bc}}.
\eeq

The which-path distinguishability $D$, i.e., the degree to which the light field at screen $c$ is coming from only one of the two slits $a$ or $b$, is given by the standard expression
\beq \label{D-expression}
D = \Big| \frac{I_{ac} - I_{bc}}{I_{ac} + I_{bc}}\Big|.
\eeq
This quantity is also called predictability by Jaeger, Horne and Shimony \cite{Jaeger-etal}.

From these we recover the well-known duality inequality (\ref{V2D2}):
\beq \label{V2D2again}
V^2 + D^2 = 1 - \frac{4{I_{ac}I_{bc}} (1 - |\gamma|^2)}{(I_{ac}+I_{bc})^2}\le1.
\eeq \\

\noindent{\bf Calling Out Oversights.} \quad Two oversights are included in the inequality (\ref{V2D2}) itself which is valid as it stands, but its usual discussion overlooks the fact that it embodies neither completeness nor exclusivity. One notes incompleteness because the Young experiment allows for the vanishing of $V$ and $D$ together while a finite signal is nevertheless present. That is, a signal is characterized by something more, in addition to $V$ and $D$, a clear indication of incompleteness. Something is missing. 

Another oversight is to fail to see that a coherence theorem for polarization $P$ exists for $V$ and $D$ as a strictly exclusive equality \cite{PCT}: 
\beq \label{P=VD}
P^2 = V^2 + D^2. 
\eeq
This is true even if one understands $P$ more generally than in the spin sense (see \cite{PCT}). However, $V$ and $D$ are still not mutually exclusive. By changing $P$ one can reduce $D$ without increasing $V$ and vice versa -- they do not need to trade tightly with each other. Again, something seems to be missing, and this is right.

Entanglement is the missing element, confirming a suggestion made 20 years ago by Knight \cite{Knight}. Quantification of coherence in our own recent optical research \cite{QLHE, QMVE, CCC, PhysScr, PCT} has exposed connections between measures of coherence. Separately, coherence has been subjected to rigorous quantification in relation to entanglement \cite{Plenio, Adesso}. The unifying aspect is the vector space foundation common to quantum states and classical wave field theories. Independently we had already demonstrated the intimate connection of two coherences, entanglement and polarization  \cite{QMVE}. They are constrained in Young-type configurations by an additional identity:
\beq \label{PC=1}
P^2 + C^2 = 1, 
\eeq
where $C$ is the entanglement measure called concurrence \cite{Wootters}. That is, $P$ and $C$ are Young-type perfect opposites. Gaining some of $P^2$ comes at the expense of losing an equal amount of $C^2$.\\  

\noindent {\bf Theoretical Conclusion.}  The completeness, and simultaneously the exclusivity, that complementarity needs is now found by the simple step of eliminating polarization between those two identities:  $P^2 = V^2 + D^2$ and $P^2 + C^2 = 1$. This yields a tight identity:
\beq \label{VDC=1}
V^2 + D^2 + C^2 = 1.
\eeq
We believe this identity fully expresses the essence of complementarity in its most common two-slit or two-path context. Importantly, in (\ref{VDC=1}), nothing is extra and nothing is missing. Still, as a relation among three observables in a Young-type experiment, it should be open to test. 

Complementarity is understood as a universal concept, applying to both classical ray-wave and quantum particle-wave dualities, and in the same way. We have independently shown that the identity-equality (\ref{VDC=1}) has exactly the same form in a one-particle quantum context as the classical wave result here \cite{Qian-etal}. To emphasize both this universality and the significance of vector-space entanglement in both domains, we have chosen to address the challenge of confirming the identity-equality by using classical light, recording observations of all three: visibility, distinguishability and entanglement.  Our setup and laboratory results are sketched below. \\

\begin{center}{\large{\bf Experimental Confirmation} }\end{center}

\noindent{\bf  General Description}\quad This section presents the details of our experiment with classical optical beams to verify the completed complementarity identity (\ref{VDC=1}). We consider a specific context by preparing a uniform temporal component with a monochromatic laser. Then the temporal function can be factored out completely and the two $\vec{\phi}$ components in (\ref{incoherent-field}) can be simply replaced with two corresponding spin states $\hat{s}_a$, $\hat{s}_b$, i.e.,

 \beq \label{experiment-field}
\vec{E}(\rp,z)=A u_a(\rp,z)\hat{s}_a + Bu_b(\rp,z)\hat{s}_b,
\eeq
where we have omitted the temporal function. The mutual coherence is now simplified as inner product of the two spin components, and is determined by their relative angle $\theta$ and phase $\xi$, i.e., $\gamma = \hat{s}_a \cdot \hat{s}_b =\cos\theta e^{i\xi}$.

\begin{figure}[h!]
\includegraphics[width=8cm]{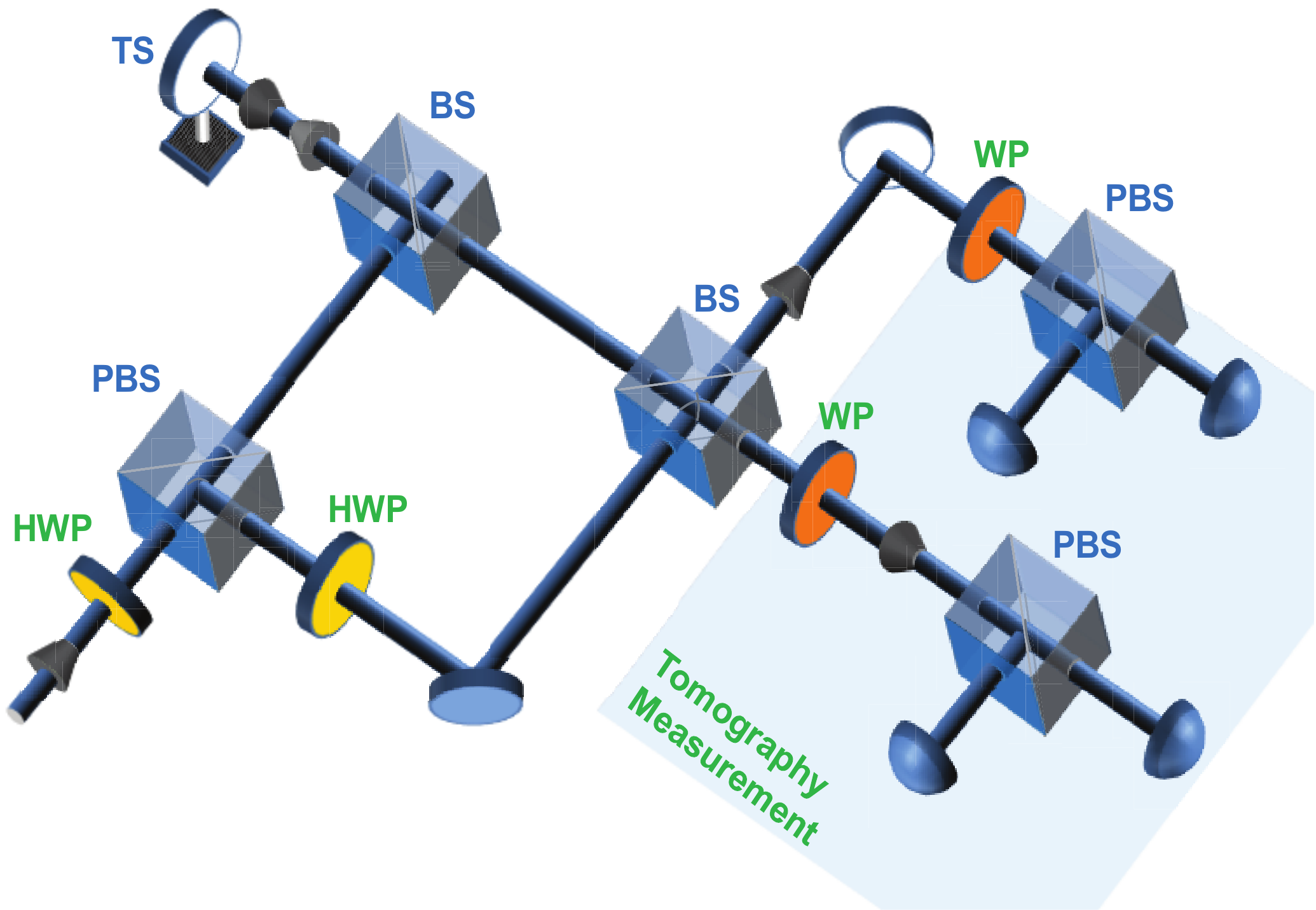}
\caption{Experimental setup. An elaborate version of Fig. \ref{MachZ} with the entanglement tomography setup shown. }
\label{MachZ2}
\end{figure}

As shown in Fig. \ref{MachZ2}, the entire scheme contains a preparation stage realized by the modified Mach-Zehnder interferometer and a measurement stage with a tomography setup. The preparation stage is designed to generate optical beams of the form given in Eq. (\ref{experiment-field}).

Specifically, a single-mode laser operating at 795nm, with linear $\hat{x}$ polarization, passes through a half-wave plate (HWP1) and changes into an arbitrarily polarized state $\hat{s}$. A polarizing beam splitter (PBS) splits the beam into two by $\hat{x}$ and $\hat{y}$ polarizations. The transmitted channel (labeled as $a$) is proportional to $u_a(\rp)\hat{s}_a$, where $\hat{s}_a\equiv \hat{x}$ and $u_a(\rp)$ is the normalized transverse spatial function in path $a$. It then passes through a 50/50 beamsplitter (BS1) that directs to a mirror mounted on a translation stage. The reflected channel (labeled $b$) of PBS passes through a half-wave plate (HWP2) to become proportional to $u_b(\rp)\hat{s}_b$ where $\hat{s}_b=e^{i\xi}\cos\theta \hat{x} + \sin\theta\hat{y}$ and $u_b(\rp)$ is the corresponding transverse spatial dependence. The two spatial functions are orthogonal to each other, i.e., $\int u^*_au_bdxdy=\delta_{ab}$ simply because they have no overlaps. Then the light field is characterized exactly by (\ref{experiment-field}). Here parameter ratio $|A/B|$ is controlled by HWP1 and $\gamma = \cos\theta e^{i\xi}$ is controlled by the combination of HWP2 in channel $b$ (in terms of $\theta$) and the translation stage in channel $a$ (in terms of phase $\xi$). The details of light beam state preparation are described in the following subsection.

The output 50/50 beamsplitter (BS2) combines channels $a$ and $b$ to produce interference. The light beam then enters the measurement stage so that $V$, $D$, and $C$ can be registered. Fringe visibility $V$ can be simply achieved by continuously registering the intensity at the output of BS2 while moving the translation stage. The systematic maximum visibility that is achievable for our MZI is $98.1\%$, obtained by producing equal intensities of the two channels and measuring interference intensities with a polarizer placed right after BS2. All measurements of visibilities in other arrangements are corrected by this systematic maximum. The which-way distinguishability $D$ can be obtained straightforwardly with intensity measurements by blocking one of the two channels $a$ or $b$. Measurement of entanglement, quantified by concurrence $C$, between the spatial degree of freedom $\{ u_a, u_b\}$ and polarization space $\{ \hat{x}, \hat{y}\}$ is realized by a tomography setup as shown in Fig. \ref{MachZ2}. It is a joint measurement of the Stokes-like parameters in both degrees of freedom \cite{James-etal}. The details of this space-polarization tomography measurements are described in the third subsection. \\

\begin{figure}[b!]
\includegraphics[width=6cm]{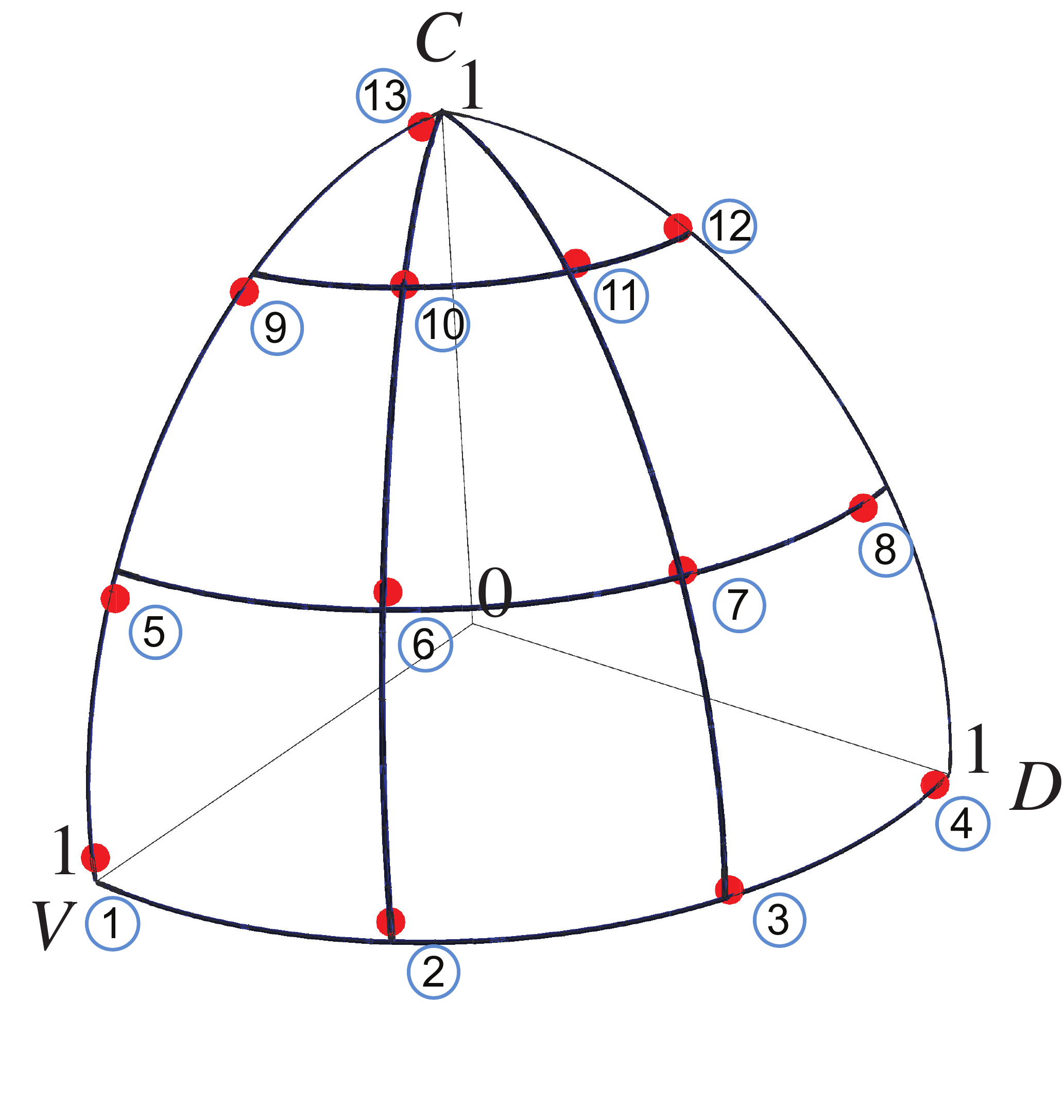}
\caption{Complementarity Sphere. Thirteen target and actually measured sets of  $(V,D,C)$ are identified by red dots.} 
\label{sphere}
\end{figure}

\noindent {\bf Beam Preparation} \quad The theoretical result (\ref{VDC=1}) predicts the values of $(V, D, C)$ to represent a spherical surface. In our experimental test, we tested 13 representative sets of $V,D,C$ distributed over an octant of this surface. These target sets are determined respectively by the nodes of seven grid lines on a Complementarity Sphere, i.e., $V=0$, $D=0$, $C=0$, $C=1/2$, $C=\sqrt{3}/2$, $V/D=\sqrt{3}$, and $V/D=1/\sqrt{3}$, shown in Fig.~\ref{sphere}. The detailed values of all sets ($V,D,C$) are given in detail below.

Now we describe how to generate these target $(V, D, C)$ values with light beams given in Eq.~(\ref{experiment-field}). The visibility and distinguishability are given in (\ref{V-expression}) and (\ref{D-expression}) respectively, and we re-express them in terms of the controllable amplitude ratio $R=|B/A|$ and $\cos\theta$ as
\beq \label{DV-parameter}
D=\left| \frac{1- R^2}{1+R^2}\right|, \quad  V=\frac{2R|\cos\theta |}{1+R^2}.
\eeq

The entanglement (concurrence) between the spatial degree of freedom $\{u_a,u_b\}$ and polarization degree of freedom $\{\hat{x},\hat{y}\}$ is given as:
\beq \label{C-generic}
C = \frac{2\sqrt{(1-|\gamma|^2)I_{ac}I_{bc}}}{I_{ac} + I_{bc}}.
\eeq
and is re-expressed as
\beq \label{C-parameter}
C=\frac{2R|\sin\theta |}{1+R^2}.
\eeq

With the above expressions (\ref{DV-parameter}) and (\ref{C-parameter}), the ratio $R$ and $\theta$ of the experimental beam (\ref{experiment-field}) can be determined for each given set of $(V,D,C)$ values. The experimentally relevant controls of $R$ and $\cos\theta$ are realized by adjusting the two half-wave plates HWP1 and HWP2 respectively.

We have prepared four sets of $(V,D,C)$ targeted at zero entanglement $C=0$, given as $(1,0,0)$ and $(\frac{\sqrt{3}}{2},\frac{1}{2},0)$ and $(\frac{1}{2},\frac{\sqrt{3}}{2},0)$ and $(0,1,0)$. These correspond to a uniform $\cos\theta=1$ and the ratios $R^2=1$ and $1/3$ and $(2-\sqrt{3})/(2+\sqrt{3})$ and $0$ respectively. Experimentally, $\cos\theta=1$, i.e., $\theta=0$ can be realized with the half-wave plate (HWP2) rotating channel $b$'s polarization to $\hat{x}$ (same as in channel $a$), and the ratio $R^2$ is continuously adjustable with the manipulation of HWP1.

The second four sets of $(V,D,C)$ are designed to have entanglement $C=1/2$, i.e., $(\frac{\sqrt{3}}{2},0,\frac{1}{2})$ and $(\frac{3}{4},\frac{\sqrt{3}}{4},\frac{1}{2})$ and $(\frac{\sqrt{3}}{4},\frac{3}{4},\frac{1}{2})$ and $(0,\frac{\sqrt{3}}{2},\frac{1}{2})$. Their corresponding $|\cos\theta |$ values are $\sqrt{3}/2$ and $3/\sqrt{13}$ and $\sqrt{3/7}$ and $0$ respectively, and the respective ratios $R^2$ are given as $1$ and $(4-\sqrt{3})/(4+\sqrt{3})$ and $1/7$ and $(2-\sqrt{3})/(2+\sqrt{3})$. The following four sets of $(V,D,C)$ have fixed entanglement $C=\sqrt{3}/2$, specifically given as $(\frac{1}{2},0,\frac{\sqrt{3}}{2})$ and $(\frac{\sqrt{3}}{4},\frac{1}{4},\frac{\sqrt{3}}{2})$ and $(\frac{1}{4},\frac{\sqrt{3}}{4},\frac{\sqrt{3}}{2})$ and $(0,\frac{1}{2},\frac{\sqrt{3}}{2})$. Their corresponding values of $|\cos\theta |$ are determined as $1/2$ and $\sqrt{1/5}$ and $\sqrt{1/13}$ and $0$ respectively, and the respective ratios $R^2$ are obtained as $1$ and $3/5$ and $(4-\sqrt{3})/(4+\sqrt{3})$ and $1/3$. Again, for these eight sets of $(V,D,C)$ the ratio $R$ and overlap parameter $\theta$ are controlled continuously with the rotations of  HWP1 and HWP2 respectively.

The final set $(V,D,C)=(0,0,1)$ corresponds to the extreme case when both $V$ and $D$ vanish. From the conventional interpretation of (\ref{V2D2again}), light has neither wave nor particle properties. But as we have pointed out, it is still a finite signal with maximal entanglement ($C=1$). In this case, $\cos\theta=0$ and $R=1$ which can be easily controlled by HWP1 and HWP2 respectively. 

The actually measured values of $V,D,C$ for their corresponding light beams are given in the main text, illustrated in Figure 2 by the red dots and presented below in detail in Table 1.\\


\begin{figure}[b!]
\includegraphics[width=3cm]{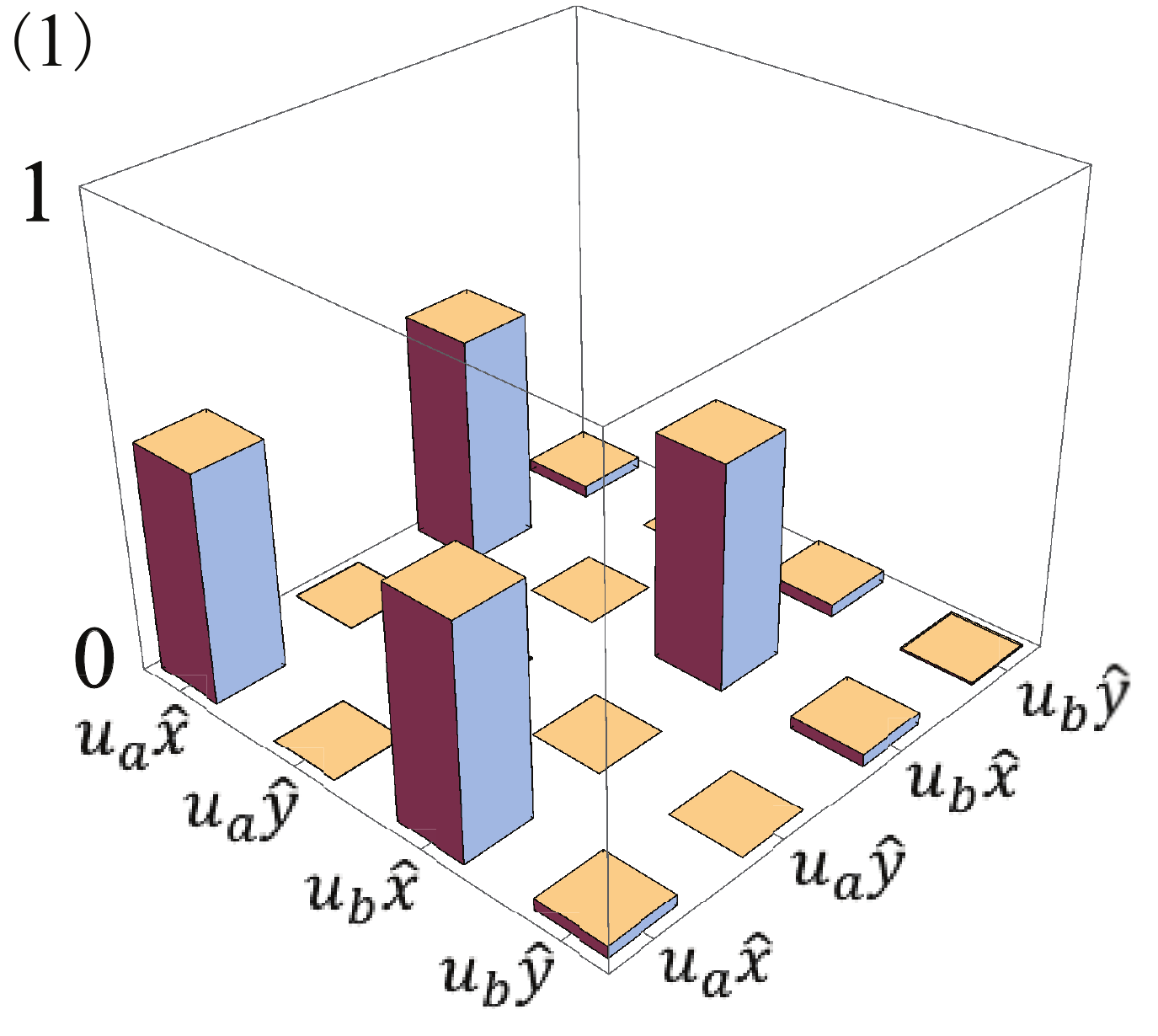}
\includegraphics[width=3cm]{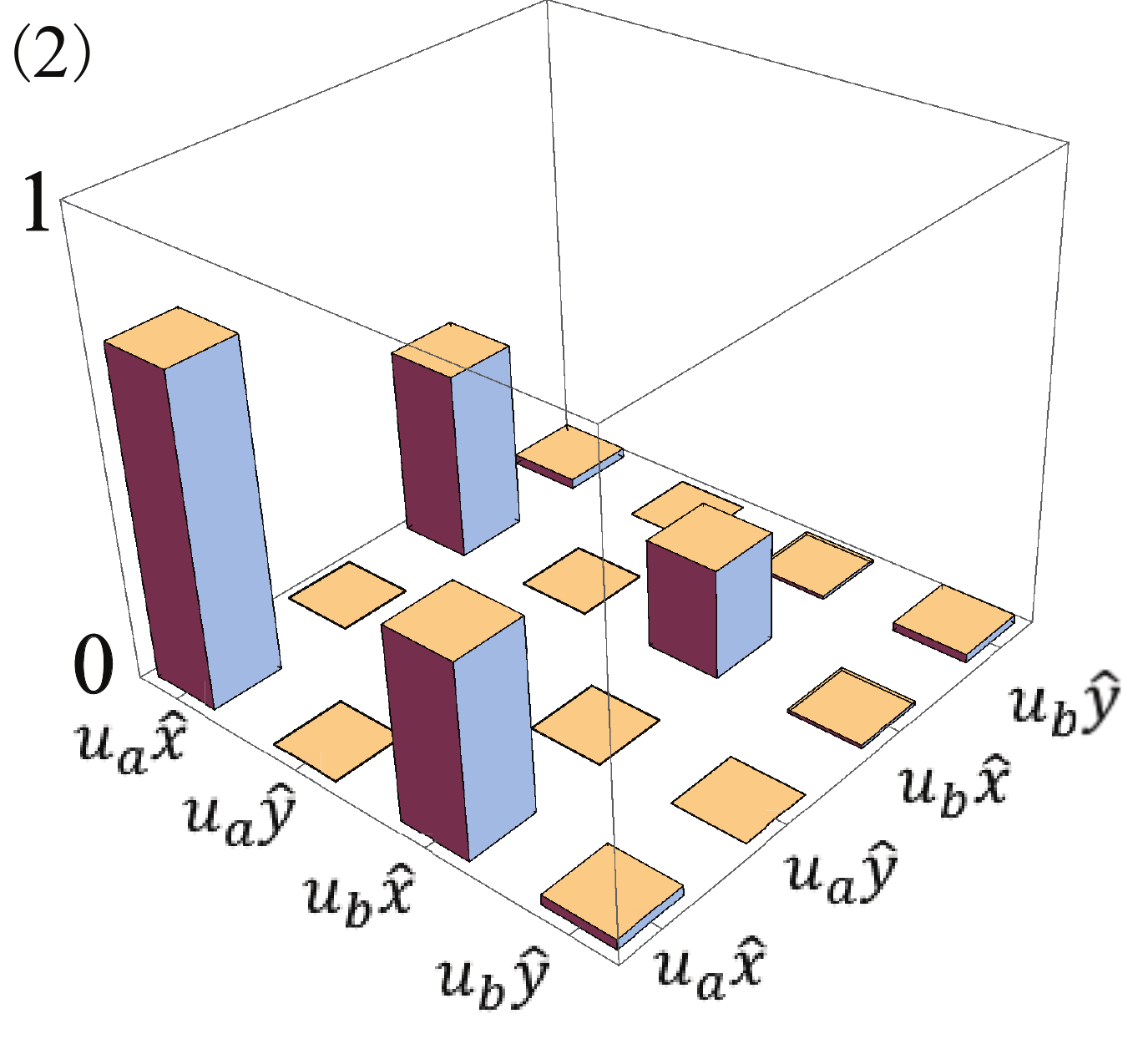}
\includegraphics[width=3cm]{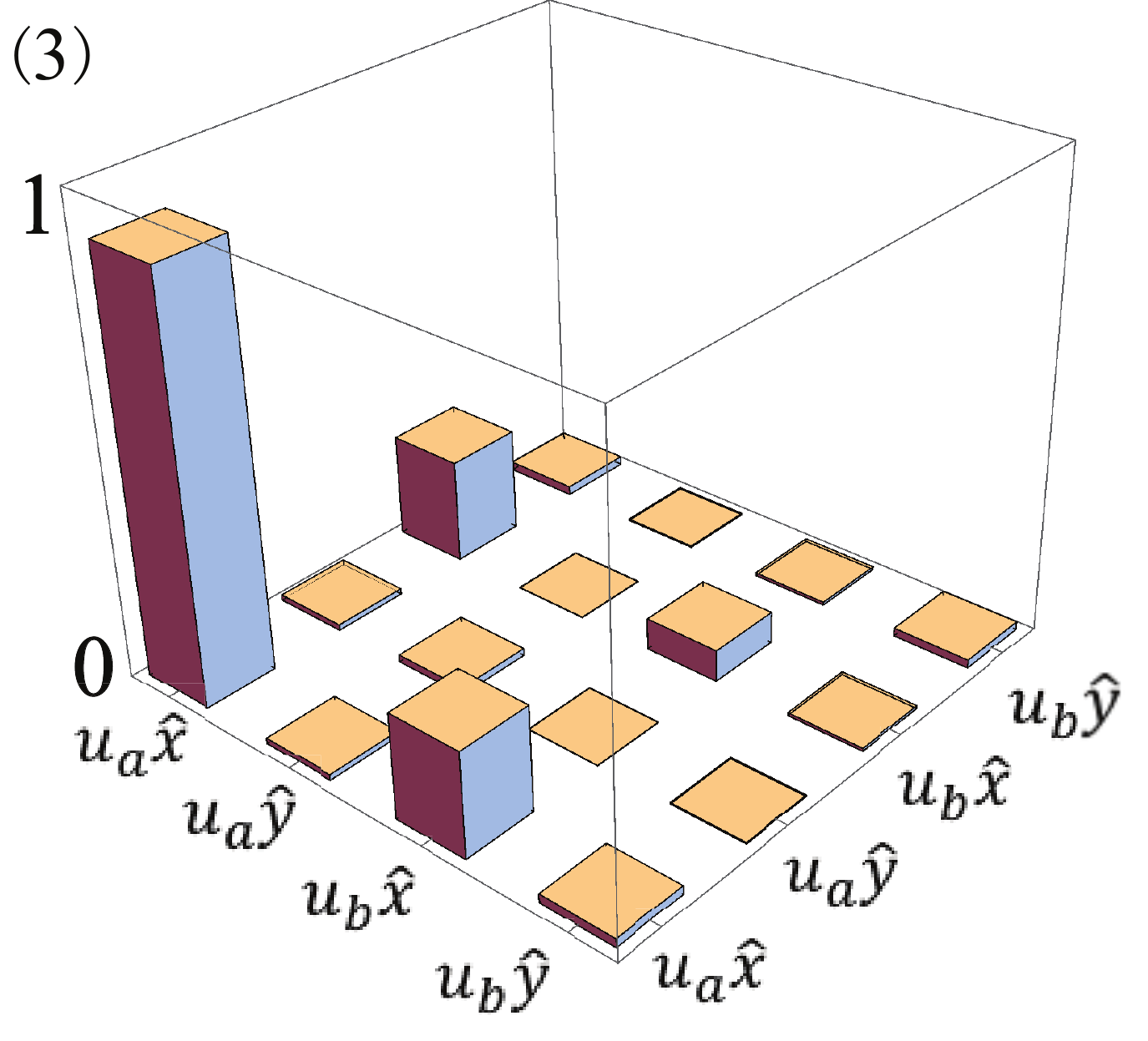}
\includegraphics[width=3cm]{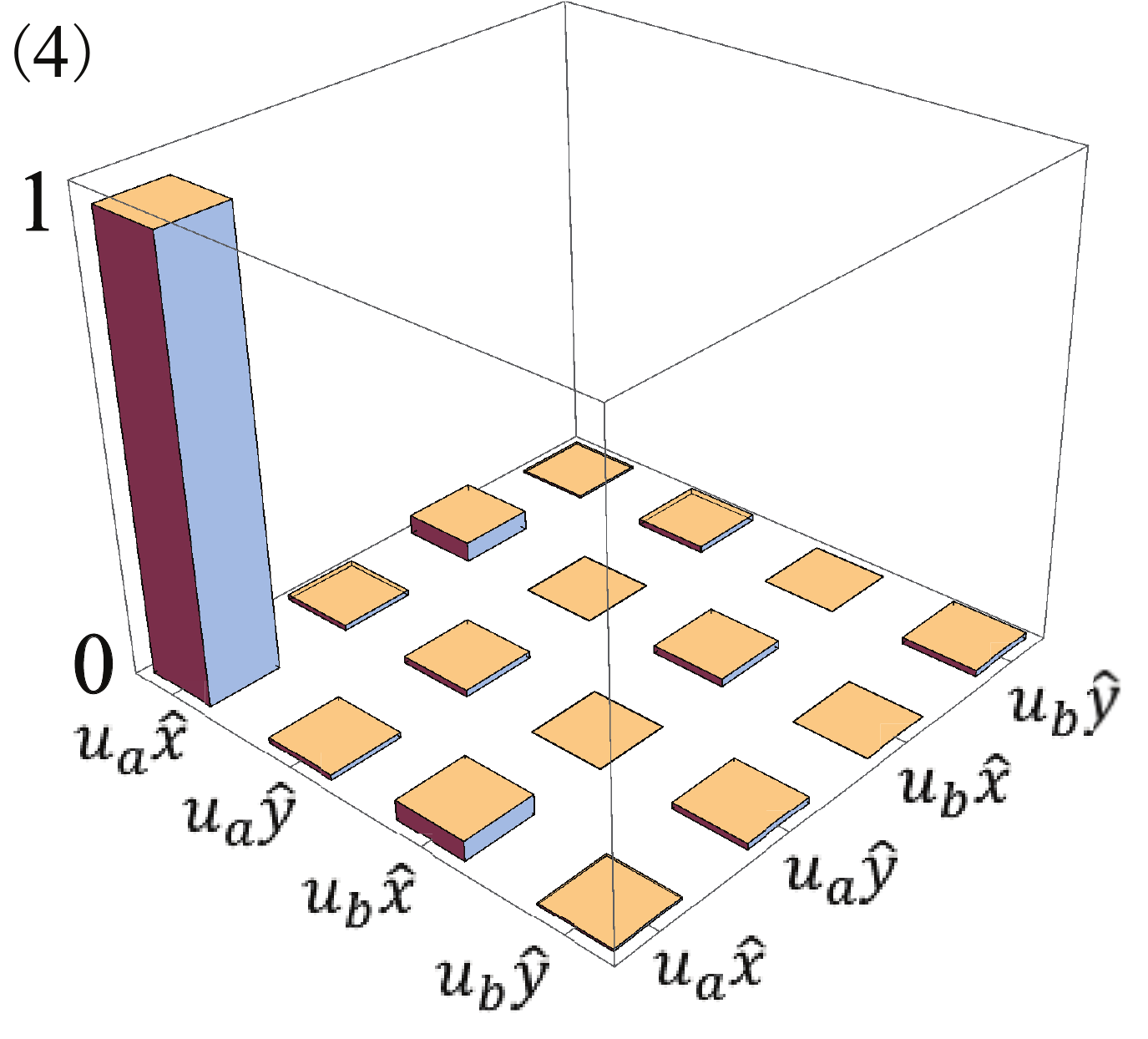}
\includegraphics[width=3cm]{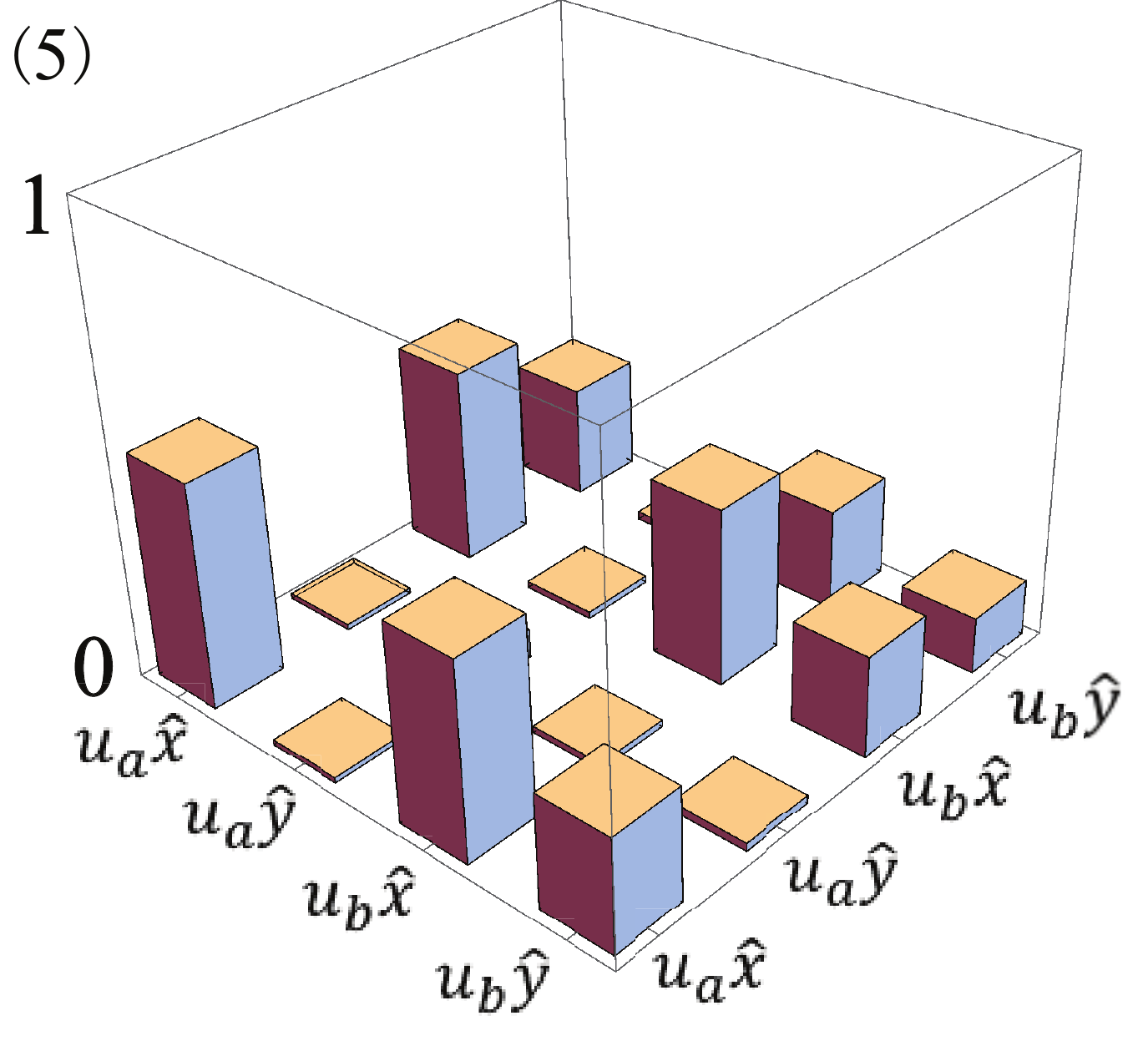}
\includegraphics[width=3cm]{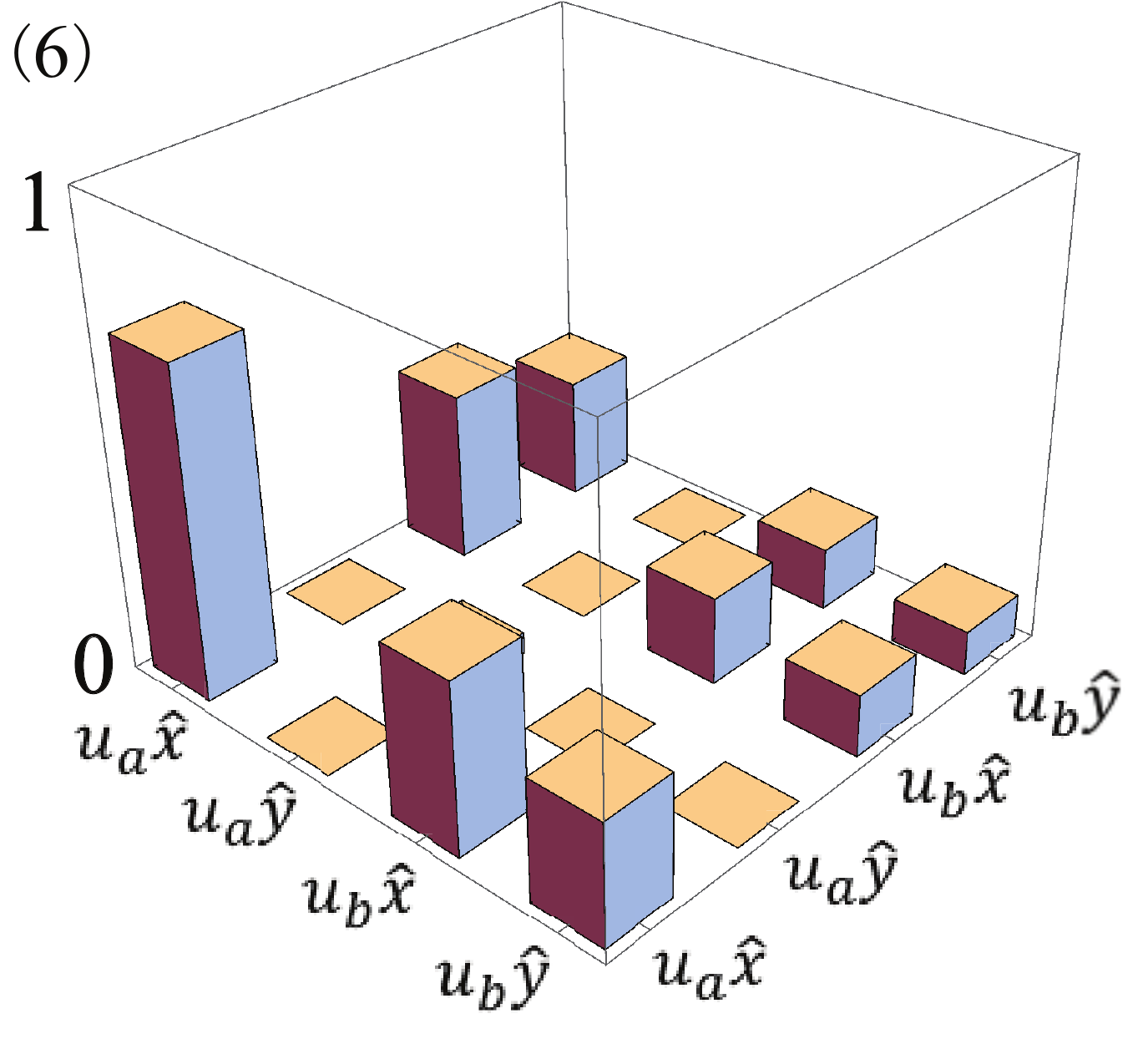}
\includegraphics[width=3cm]{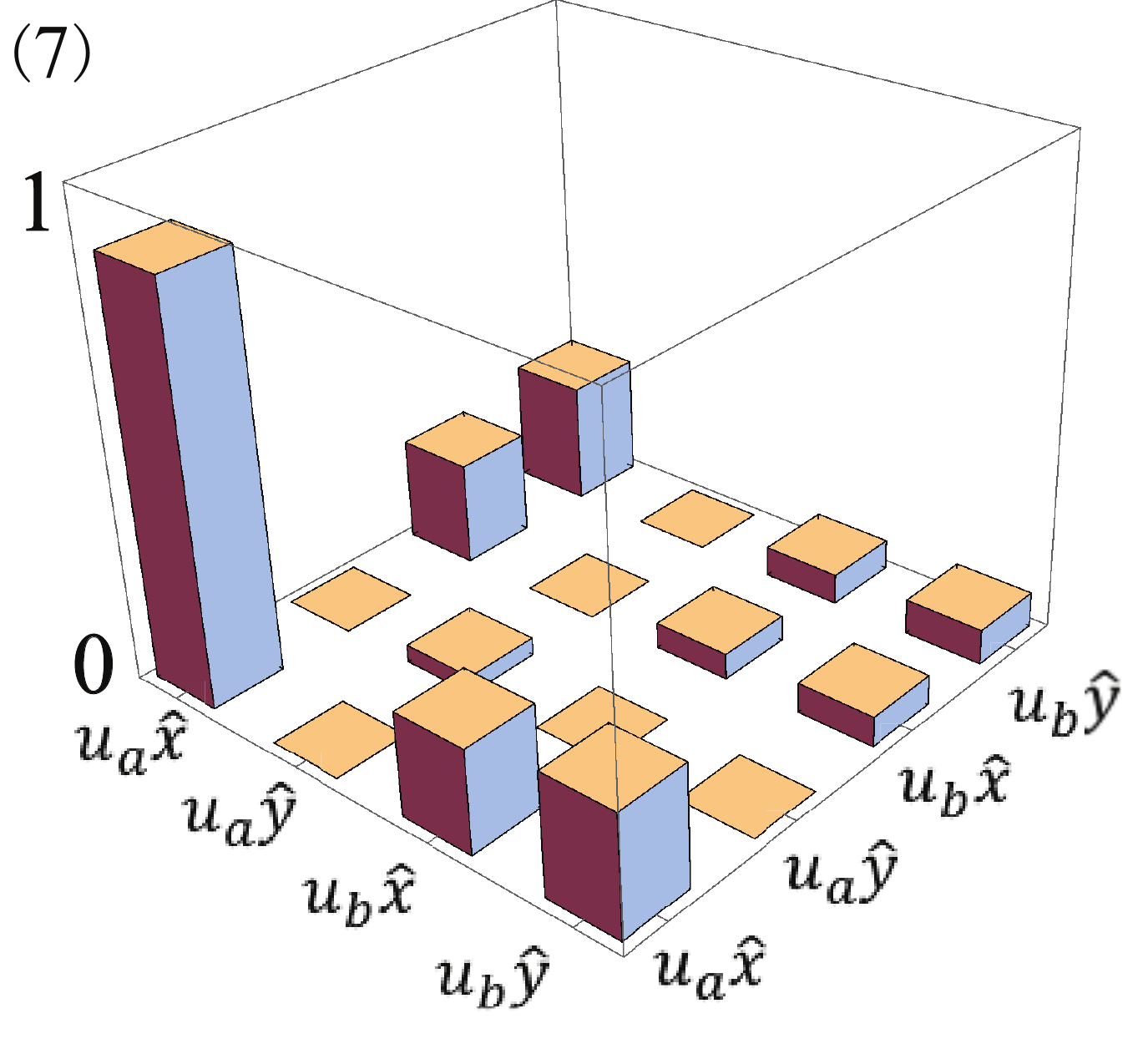}
\includegraphics[width=3cm]{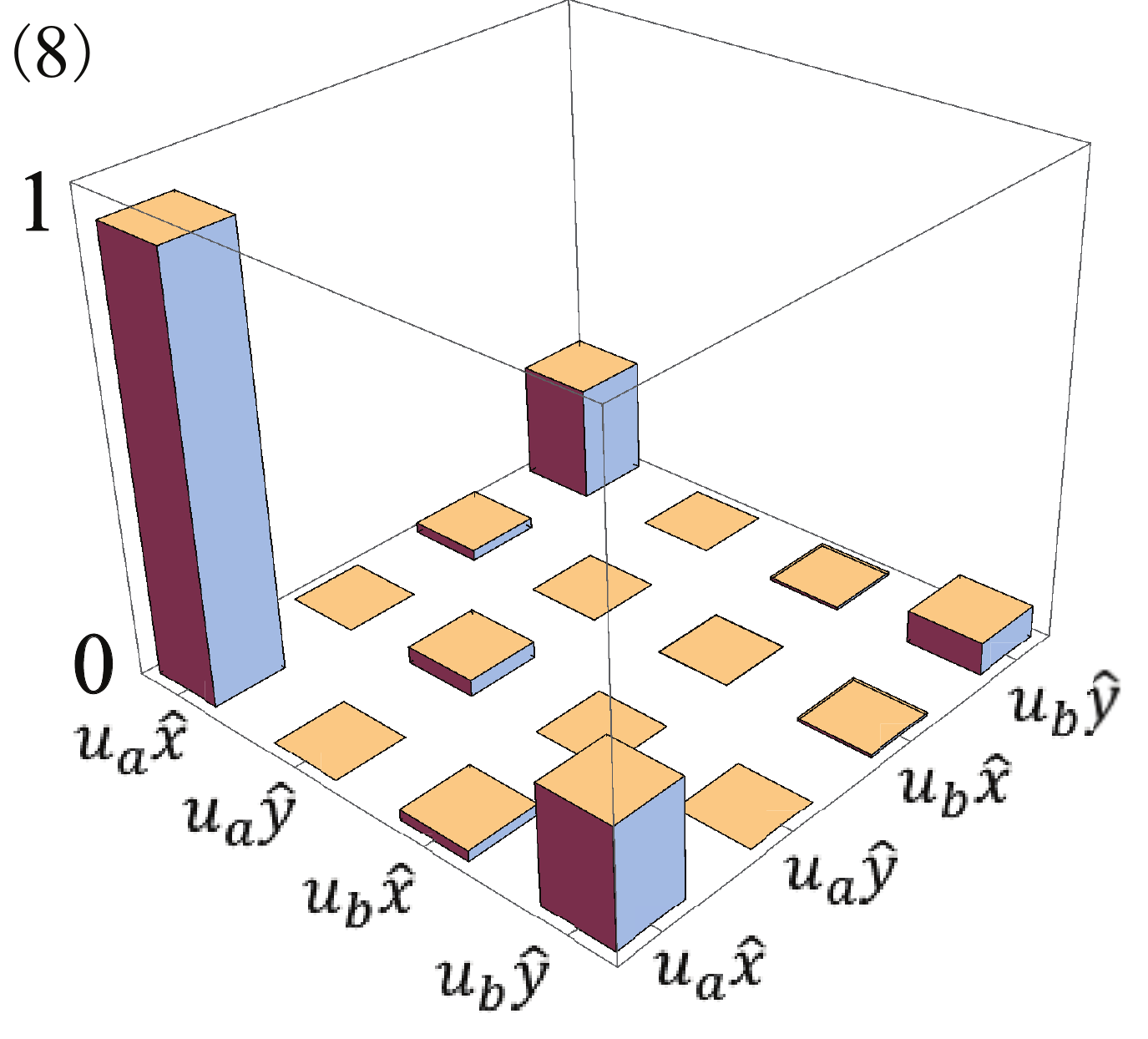}
\includegraphics[width=3cm]{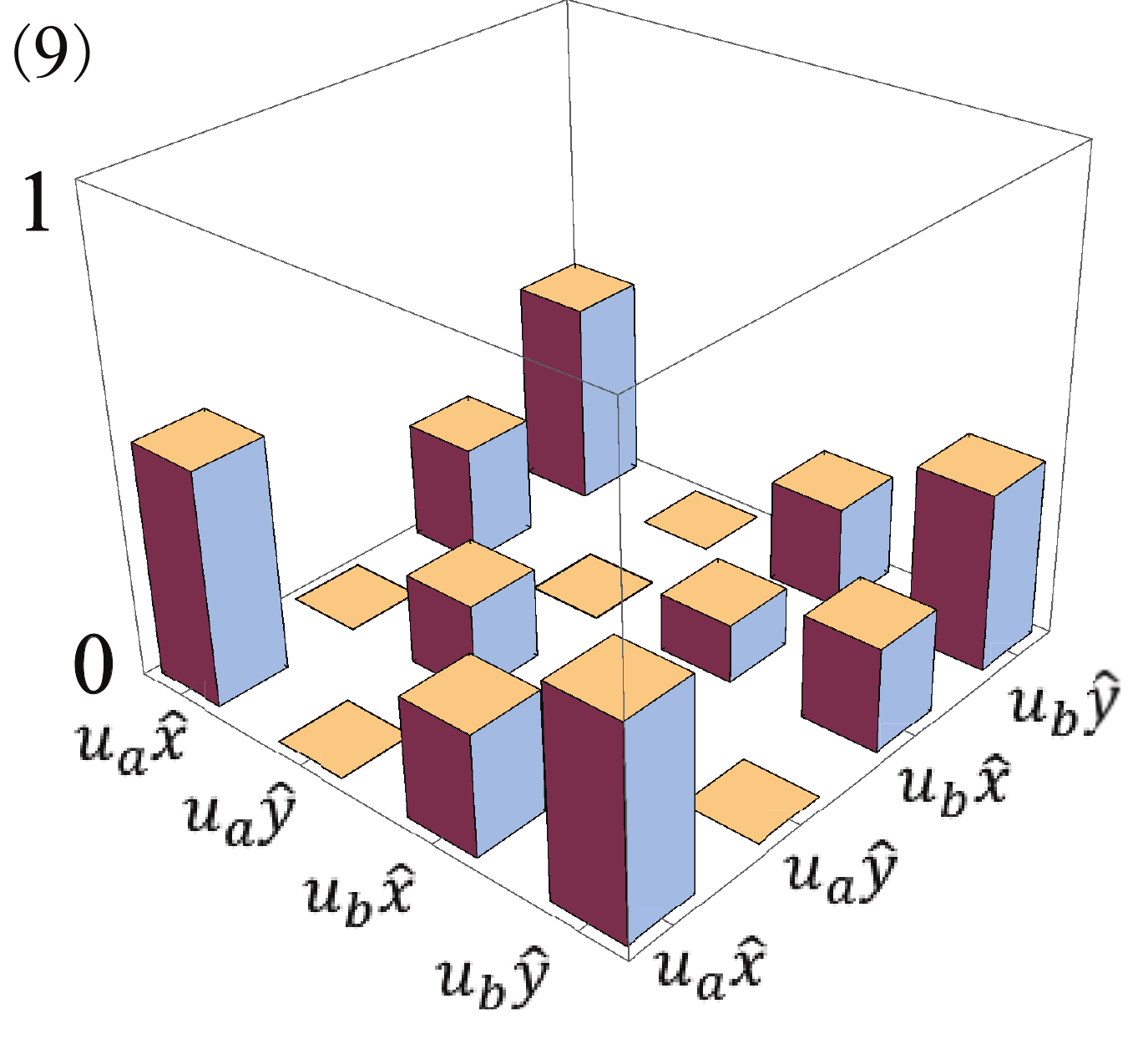}
\includegraphics[width=3cm]{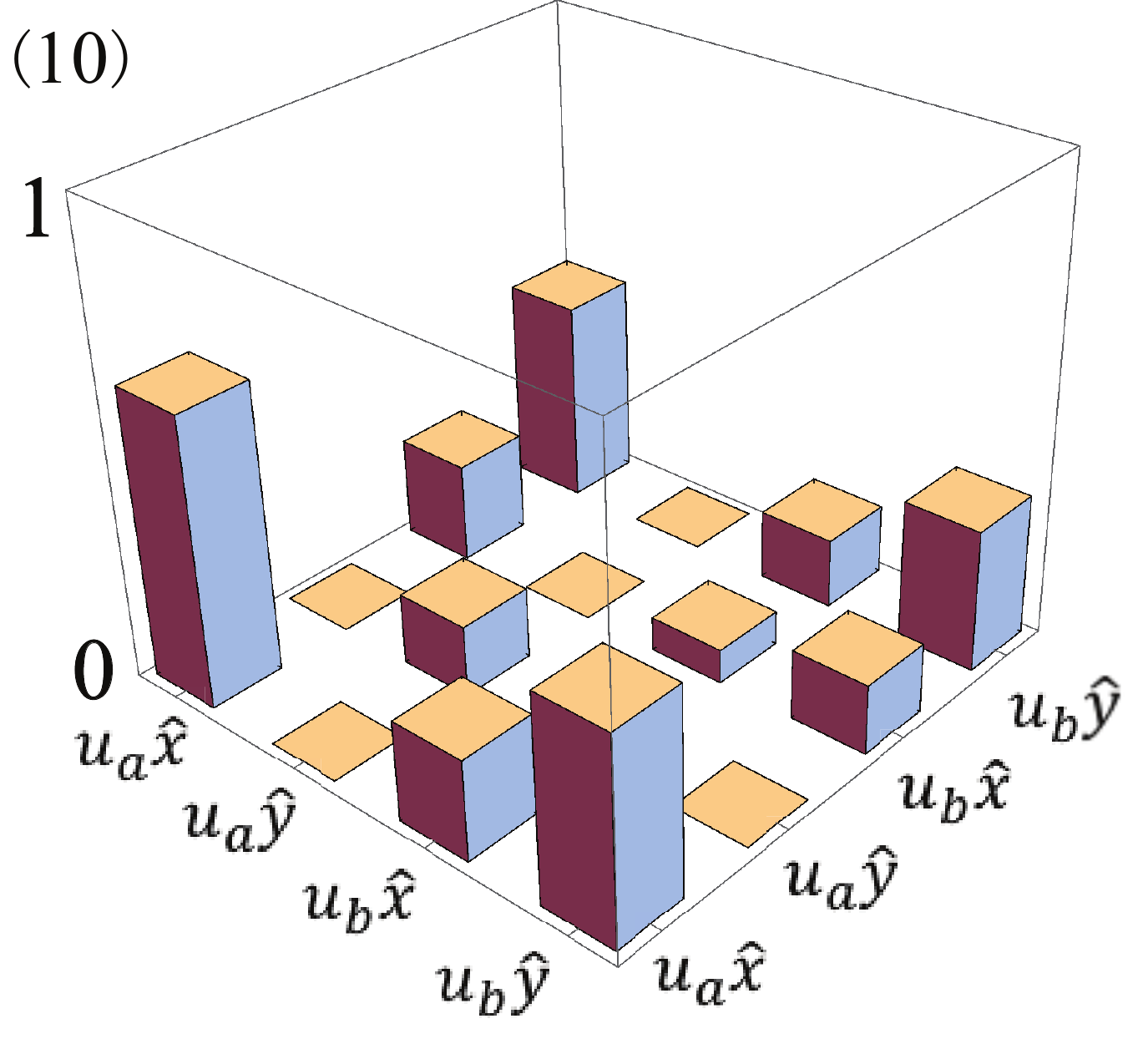}
\includegraphics[width=3cm]{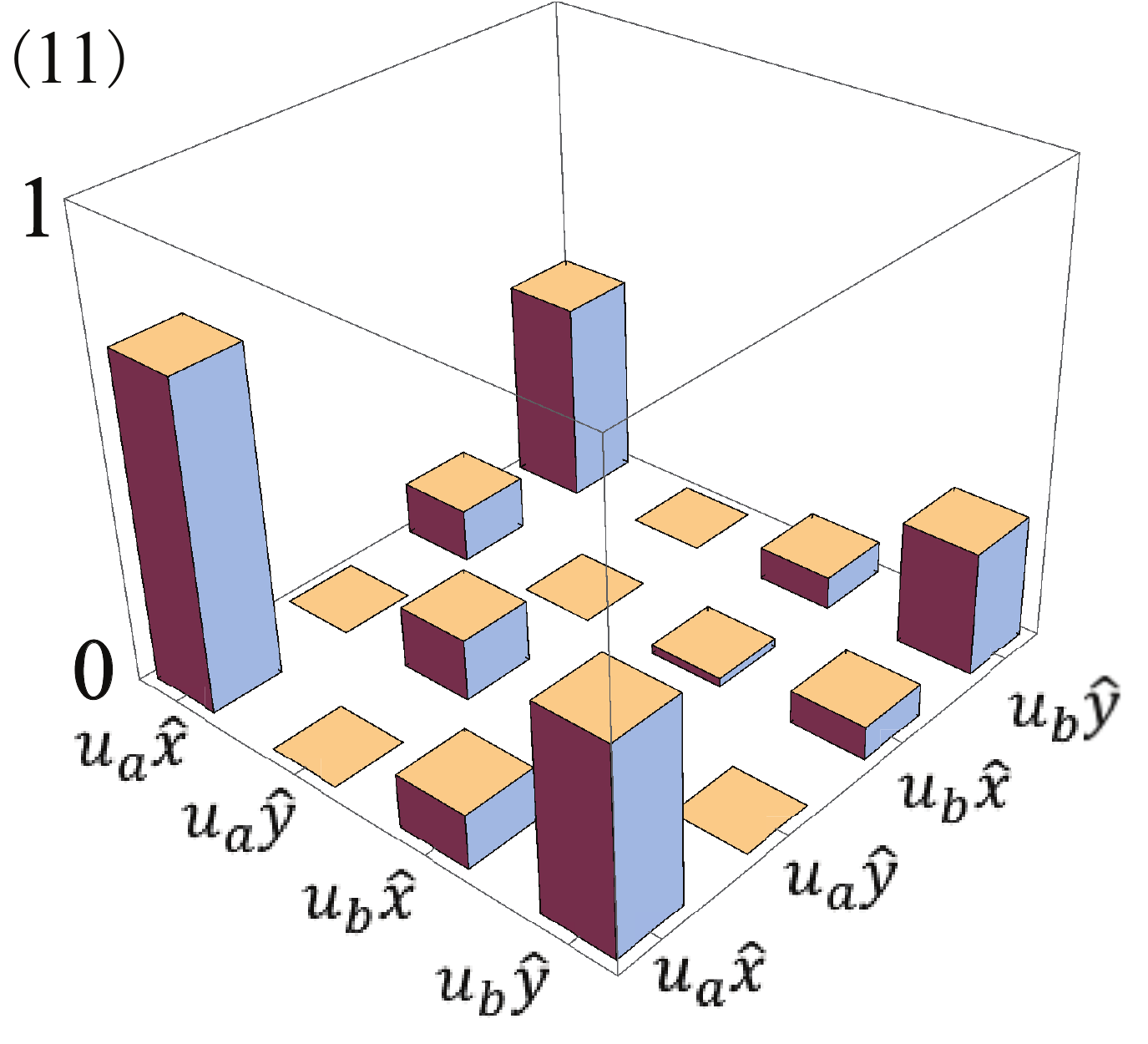}
\includegraphics[width=3cm]{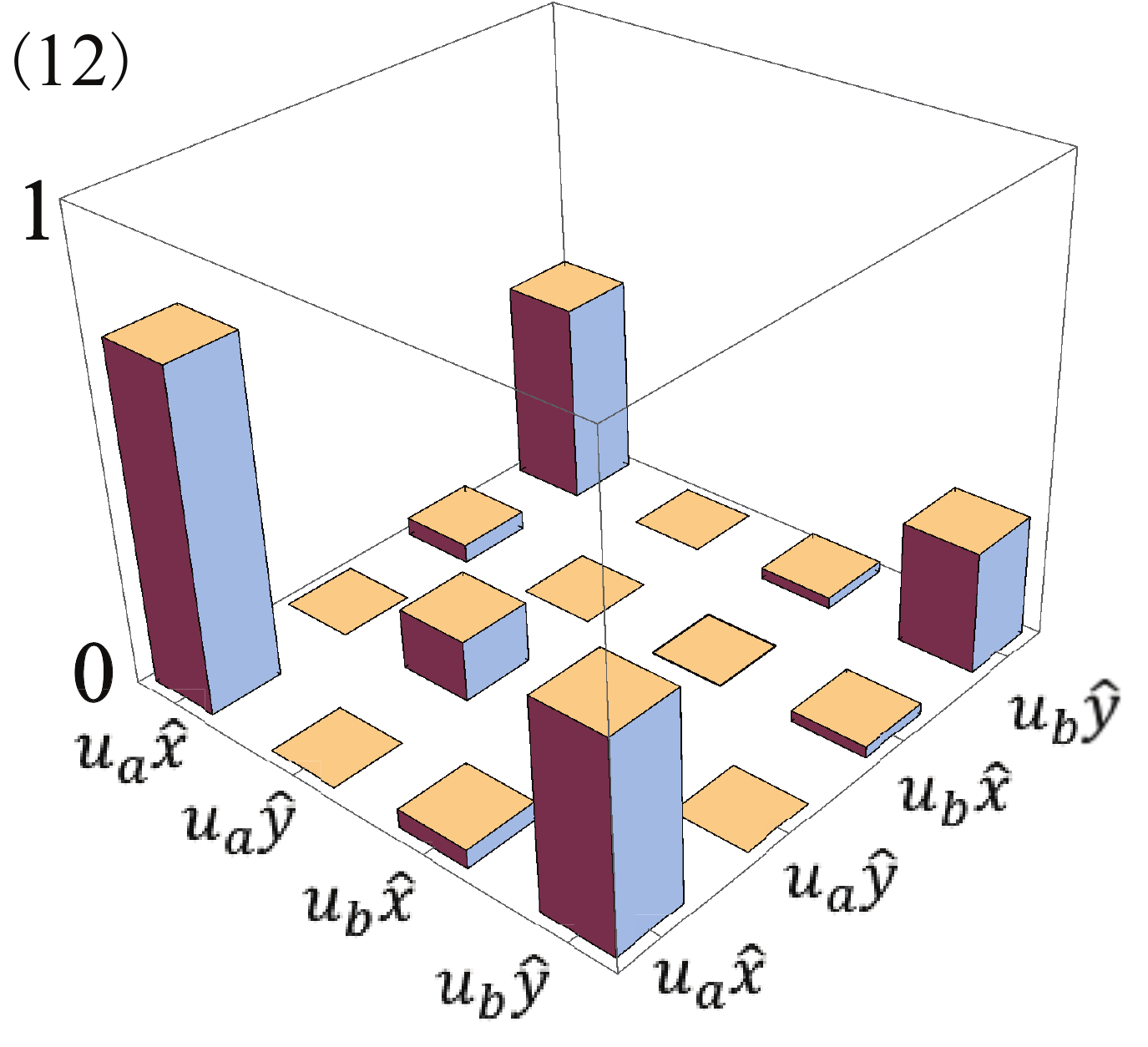}
\includegraphics[width=3cm]{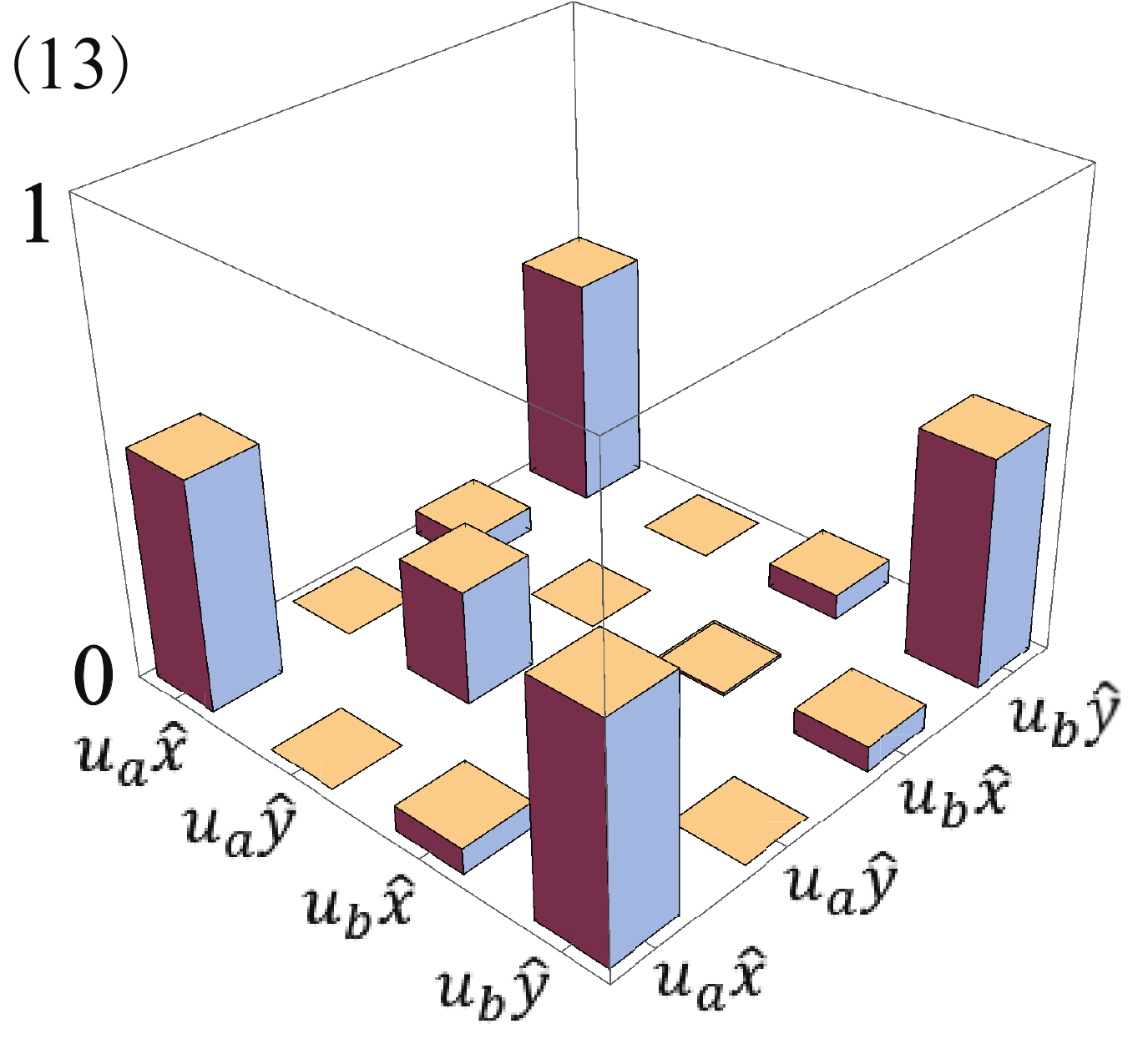}
\caption{Tomographic coherence matrices of the joint spatial and polarization degrees of freedom for thirteen representative beams.} 
\label{matrices}
\end{figure}

\noindent {\bf Tomography Measurement} \quad Here we describe the details of our spatial-spin two degrees of freedom tomography measurements. Each degree of freedom has two effective dimensions, i.e., $u_a$, $u_b$ in the spatial domain and $\hat{x}$, $\hat{y}$ in the spin polarization domain. Ideally, the light beam should be characterized by the temporally coherent expression (\ref{experiment-field}). However, in reality the temporal amplitude of the two paths may not be perfectly coherent, or the contents of the modes are not perfectly pure. These cases result in a matrix description of the light beam in the basis $\{u_a\hat{x},u_a\hat{y},u_b\hat{x},u_b\hat{y}\}$, given as
\begin{eqnarray} \label{cohmatrix}
{\cal W} = \left[\begin{matrix}
{\cal W}_{1,1}  & {\cal W}_{1,2} &{\cal W}_{1,3}&{\cal W}_{1,4}\\
{\cal W}_{2,1}  & {\cal W}_{2,2} &{\cal W}_{2,3}&{\cal W}_{2,4}\\
{\cal W}_{3,1}  & {\cal W}_{3,2} &{\cal W}_{3,3}&{\cal W}_{3,4}\\
{\cal W}_{4,1}  & {\cal W}_{4,2} &{\cal W}_{4,3}&{\cal W}_{4,4}
\end{matrix} \right].
\end{eqnarray}
One can call this a coherence matrix for two degrees of freedom \cite{Abouraddy-etal}. 

Then, following the standard two-qubit tomography procedure \cite{James-etal}, the output beams are measured jointly in the Stokes-like basis of both degrees of freedom to recover each matrix element in ({\ref{cohmatrix}}). In principle, one needs to project the light onto the following 36 joint bases, i.e., $\{u_a, u_b, (u_a\pm u_b)/\sqrt{2}, (u_a\pm i u_b)/\sqrt{2}\}\otimes\{\hat{x}, \hat{y}, (\hat{x}\pm \hat{y})/\sqrt{2}, (\hat{x}\pm i \hat{y})/\sqrt{2}\}$, six bases in each degree of freedom. The projections of light in the six spin polarization state bases are standard by the usage of quarter-wave and half-wave plates combining with polarizing beam splitters as shown in the tomography measurement section of the setup in Fig. \ref{MachZ2}.

The projection in the spatial basis $\{u_a, u_b\}$ is realized straightforwardly by blocking one of the two channels of the Mach-Zehnder interferometer. The projection onto the bases $(u_a\pm u_b)/\sqrt{2}$ can be realized by leaving both channels open. The 50/50 beamsplitter (BS2) is an unitary rotation operator of the two incoming paths. It operates as an effective projection with one output port in the basis $(u_a + u_b)/\sqrt{2}$ and the other port in $(u_a - u_b)/\sqrt{2}$. Finally, the projection in the basis $(u_a\pm i u_b)/\sqrt{2}$ is achieved with the combination of the 50/50 beamsplitter BS2 and the translation stage that can introduce an additional $\pi/2$ phase in path $a$.

With the above arrangements, one is able to realize measurements in all joint bases. Then the coherence matrices ({\ref{cohmatrix}}) for all 13 beams can be reconstructed in the standard basis $u_a\hat{x}, u_a\hat{y}, u_b \hat{x}, u_b\hat{y}$. The actually measured matrix elements are shown respectively in Fig.~\ref{matrices}.\\

\noindent {\bf Confirmation} \quad The completed complementarity identity (\ref{VDC=1}) calls to mind a sphere in a space with independent $V$-$D$-$C$ axes. Our measurements of the $VDC$ variables for different field states are identified on a sphere (one octant is enough) by thirteen points indicating individual measurements, as in Fig. \ref{sphere}. \\

The measured $VDC$ values for each of those points, as well as the sum $V^2 + D^2 + C^2 $, are shown in Table \ref{Table} below. From the table, one concludes that the three-way complementarity identity (\ref{VDC=1}) connecting entanglement-distinguishability-visibility is well confirmed.\\

\begin{table}[t!] 
\begin{center}
\begin{tabular}{|r|r|r|r|r|l|l|r|ll}
\hline
&  $V\ $ &$D\ $&$C\ $& ${\rm SUM\ }$ \\
\hline 
\hspace{0.5mm}
State 1\ & \ \ 0.996 & \ \ 0.004 & \ \ 0.046 &\ \ 0.993\\
\hline 
\hspace{0.5mm}
State 2\ & 0.863  &0.498&0.035&0.994 \\
\hline 
\hspace{0.5mm}
State 3\ & $0.488$ &0.867&0.009&0.990  \\
\hline 
\hspace{0.5mm}
State 4\ & 0.046  & 0.996 & 0.003 & 0.994\\
\hline 
\hspace{0.5mm}
State 5\ & 0.885 & 0.010 &0.463 & 0.997 \\
\hline 
\hspace{0.5mm}
State 6\ & 0.733 & 0.431 &0.522 &0.996 \\
\hline 
\hspace{0.5mm}
State 7\ & 0.424 & 0.747 &0.505 &0.992 \\
\hline 
\hspace{0.5mm}
State 8\ & 0.078 & 0.865 &0.494 &0.999 \\
\hline 
\hspace{0.5mm}
State 9\ & 0.528 & 0.004 & 0.845 &0.993 \\
\hline 
\hspace{0.5mm}
State 10\ & 0.426 & 0.247 & 0.868 &0.996 \\
\hline 
\hspace{0.5mm}
State 11\ & 0.226 & 0.430 & 0.872 &0.996 \\
\hline 
\hspace{0.5mm}
State 12\ & 0.014 & 0.484 & 0.875 & 1.000 \\
\hline 
\hspace{0.5mm}
State 13\ & 0.064 & 0.005 & 0.991 &0.987 \\
\hline 
\end{tabular}
\end{center}
\caption{The measured values of visibility, distinguishability, and entanglement, where ${\rm SUM}$ refers to the identity combination $ V^2 + D^2 + C^2 $. The maximum standard deviation of the parameter measurements is $0.042$.} \label{Table}
\end{table}

The thirteen points on the octant were obtained with a modified tomographic setup as shown in Fig. \ref{MachZ2}. The Mach-Zehnder interferometer, operated as a double slit, produces interference of two light components with visibility achieved through motion of the translation stage. Which-way distinguishability is obtained by detecting each single component, and intrinsic entanglement is determined by the follow-up tomography. The entire setup is suitable to test the completed complementarity relation (\ref{VDC=1}) for both single photon (quantum) and classical optical beam cases. \\

\begin{center}{\large{\bf Overview and Comments} }\end{center}

\noindent{\bf Overview} \quad The critical result is of course the derivation of the identity $V^2 + D^2 + C^2 = 1$ along with its experimental validation. As we pointed out, the well-known inequality (\ref{V2D2again}) doesn't embody completeness or exclusivity, the two essential elements of complementarity. Something must be missing, and our analysis is exposed more completely here. There are three useful extreme situations. 

Case (a), consider $V=1$. Then equation (\ref{V-expression}) requires $|A|=|B|$, meaning $D=0$, and $|\gamma| = 1$ (indicating $\vec{\phi}_a = e^{i\eta}\vec{\phi}_b =\vec{\phi}$). In this extreme case the field (\ref{incoherent-field}) becomes simply 
\beq \label{separable1}
\vec{E}(\rp,z,t)\propto \vec{\phi}(t)[u_a(\rp,z)+ e^{i\Theta} u_b(\rp,z)],
\eeq
where $\Theta = \eta + \arg(A^*B)$, showing $\vec{E}$ completely factorized between the spatial degree of freedom spanned by $\{u_a,u_b\}$ and the remaining degrees of freedom represented by $\vec{\phi}(t)$.

Case (b), consider $D=1$. Then equation (\ref{D-expression})  requires $|A|=0$ or $|B|=0$ (either one meaning $V=0$), and the field (\ref{incoherent-field}) becomes, e.g.,
\beq \label{separable2}
\vec{E}(\rp,z,t)\propto Au_a(\rp,z)\vec{\phi}_a(t),
\eeq
which clearly displays complete separability between the spatial and the other degrees of freedom. 

Case (c), consider $V=0$ and $D=0$. Then equations (\ref{V-expression}) and (\ref{D-expression}) require that $|A|=|B|$ and $|\gamma|=|\la\vec{\phi}^*_a\cdot \vec{\phi}_b\ra|=0$. The field then becomes 
\beq \label{maximal entangled}
\vec{E}(\rp,z,t)\propto u_a(\rp,z)\vec{\phi}_a(t) + u_b(\rp,z)\vec{\phi}_b(t),
\eeq
which is certainly non-separable. It is the classical vector space equivalent of a quantum Bell state with maximal entanglement. 

These three extreme cases indicate mutual exclusivity among three key properties: visibility, distinguishability and entanglement. We have supplied the missing measure for entanglement, namely concurrence $C$. After dividing by $\sqrt{I_{ac} + I_{bc}}$ to normalize the field in (\ref{incoherent-field}) to unity at the screen, $\vec E$ is evidently a vector-space pure state, a superposition of independent two-state products, so we can immediately apply the concurrence  formula \cite{Wootters}. For the generic field (\ref{incoherent-field}) we find
\beq \label{C-gen}
C = \frac{2\sqrt{(1-|\gamma|^2)I_{ac}I_{bc}}}{I_{ac} + I_{bc}}.
\eeq
As indicated in the text, we are rewarded with a perfect three-way equality, the identity
\beq \label{CDVc}
V^2 + D^2 + C^2  \equiv 1.
\eeq 
This completed the derivation of the main result (\ref{VDC=1}) for classical optical fields.  \\

\noindent {\bf Comments} \quad The result is universal in the sense that it is a vector space identity. It automatically applies to other physical entities, both classical and quantum, having vector space structures similar to the optical field discussed. This can be regarded as the completion of Bohr's long and fruitless search for a compact statement that persuasively summarizes complementarity. He deserves all the credit for intuitively identifying exclusivity and completeness as the essential ingredients of any universal statement, and when included in the unique way we have described, these are the keys to direct understanding. His frustration, in his rephrasings and rewritings, was unfortunately inevitable. He didn't ``just miss" the simple formula (\ref{VDC=1}). It was not available to him. The understanding that the trading of wave and particle aspects with each other can be reduced to calculation, leading to (\ref{V2D2}) and then (\ref{P=VD}), was not appreciated before the much later work of Wootters and Zurek \cite{W and Z}. Furthermore, a fully developed examination \cite{QMVE, PCT, ALuis} of polarization coherence (both ordinary and generalized) relative to entanglement (both quantum and classical), as in (\ref{PC=1}), has only much more recently been attempted, and was not known and certainly not known in a classical wave context, during Bohr's working life. Finally, although we believe that $V^2 + D^2 + C^2 = 1$ represents the closing of a 90-year discussion, we don't conclude that physics is finished with complementarity. Recent work shows that it can also be formulated for more than one entity. The first steps, that are already taken, open a window onto multi-entity coherences \cite{JSV, JB, Khoury} that are not yet well explored or even conventionally named. \\


\noindent {\bf Acknowledgements: } 
We are pleased to acknowledge communication with colleagues Iwo Bialynicki-Birula, Paul Brumer, Steven Cundiff, Justin Dressel, Berge Englert, Andrew Jordan, Peter Knight, Peter Milonni, and Michael Raymer and financial support from ARO W911NF-16-1-0162, ONR N00014-14-1-0260, and NSF grants PHY-1203931, PHY-1505189, and INSPIRE PHY-1539859.
\\


\begin{thebibliography}{99}

\bibitem{Bohr-Nature} Bohr's multiply revised and delayed report from Como appeared as ``The Quantum Postulate and the Recent Development of Atomic Theory," \nat{121}, 580 (1928), and see Bohr's book {\em Atomic Theory and the Description of Nature}, (Cambridge Univ. Press, Cambridge, 1934), p. 10.

\bibitem{Jammer1} See M. Jammer for an account of subsequent discussions engaging Born, Fermi, Heisenberg, Kramers, Pauli, Rosenfeld, von Neumann and Wigner in {\em The Conceptual Development of Quantum Mechanics} (McGraw-Hill, New York, 1966), p. 354.

\bibitem{Murdoch} D. Murdoch, {\em Niels Bohr's Philosophy of Physics} (Cambridge Univ. Press, 1987).

\bibitem{Whitaker} A. Whitaker, {\em Einstein, Bohr and the Quantum Dilemma}, 2nd Edition (Cambridge Univ. Press, 2006).

\bibitem{SciAm} See B.-G. Englert, M.O. Scully and H. Walther ``The Duality in Matter and Light", Sci. Am. {\bf 271}, 86 (1994).

\bibitem{NatNat} B.-G. Englert, M.O. Scully and H. Walther, ``Complementarity and Uncertainty", \nat{375}, 367-368 (1995); and S. Durr, T. Nonn, and G. Rempe, ``Origin of quantum-mechanical complementarity probed by a 'which-way' experiment in an atom interferometer," \nat{395}, 33-37 (1998). 

\bibitem{Englert} B.-G. Englert, ``Fringe Visibility and Which-Way Information: An Inequality," \prl{77}, 2154 (1996).

\bibitem{Menzel} R. Menzel, D. Puhlmann, A. Heuer and W.P. Schleich, ``Wave-particle dualism and complementarity unraveled by a different mode", PNAS {\bf 109}, 9314 (2012).

\bibitem{DeZela} See F. De Zela, ``Relationship between the degree of polarization, indistinguishability, and entanglement", \pra{89}, 013845 (2014) as well as  M. Jammer, {\em The Philosophy of Quantum Mechanics} (Wiley \& Sons, New York, 1974).

\bibitem{Abouraddy} K.H. Kagalwala, G. di Giuseppe, A.F. Abouraddy and B.E.A. Saleh,  \npho {7}, 72 (2013).


\bibitem{PhysScr} J.H. Eberly, Xiao-Feng Qian, Asma Al Qasimi, Hazrat Ali, M. A. Alonso, R. GutiŽrrez-Cuevas, Bethany J. Little, John C. Howell, Tanya Malhotra and A.N. Vamivakas, ``Quantum and classical optics -Ð emerging links", Phys. Script. {\bf 91}, 063003 (2016).

\bibitem{QMVE} X.-F. Qian, T. Malhotra, A.N. Vamivakas and J.H. Eberly, ``Coherence Constraints and the Last Hidden Optical Coherence", \prl{117}, 153901 (2016).

\bibitem{PCT} J.H. Eberly, X.-F. Qian and A.N. Vamivakas, ``Polarization Coherence Theorem", Optica {\bf 4}, 1113 (2017). 

\bibitem{Plenio} T. Baumgratz, M. Cramer, and M. B. Plenio, ``Quantifying Coherence" \prl{113}, 140401 (2014).

\bibitem{Adesso} A. Streltsov, U. Singh,  H. S. Dhar, M. N. Bera, and G. Adesso, ``Measuring Quantum Coherence with Entanglement" \prl{115}, 020403 (2015).

\bibitem{JSV} G. Jaeger, A. Shimony and L. Vaidman, ``Two interferometric complementarities", \pra{51}, 54 (1995). See comment in Supplementary Note 3.

\bibitem{BEG-JAB} B.G. Englert and J.A. Bergou, ``Quantitative Quantum Erasure", \oc{179}, 337 (2000).

\bibitem{ALuis} A. Luis, ``Coherence, polarization and entanglement for classical light fields", \oc{282}, 3665 (2009). See comment in Supplementary Note 3.

\bibitem{JB} M. Jakob and J.A. Bergou, ``Quantitative complementarity relations in bipartite systems: Entanglement as a physical reality", \oc{283}, 827 (2010). See comment in Supplementary Note 3.

\bibitem{Khoury} W. F. Balthazar, C. E. R. Souza, D. P. Caetano, E. F. Galv\~ao, J. A. O. Huguenin, and A. Z. Khoury, ``Tripartite nonseparability in classical optics," \ol{41}, 5797 (2016). See comment in Supplementary Note 3.

\bibitem{names} Attendees at Como included such famous names in physics as: Born, Bose, Bragg, Brillouin, de Broglie, Compton, Debye, Fermi, Franck, Frenkel, Gerlach, Heisenberg, von Laue, Lorentz, Majorana, Millikan, von Neumann, Paschen, Pauli, Planck, Rutherford, Sommerfeld, Stern, Tolman, Wigner, Wood, and Zeeman. Apparently Ehrenfest and Einstein were invited but did not attend. Schr\"odinger's absence is puzzling.

\bibitem{Schilpp} P.A. Schilpp, ed., {\em Albert Einstein: Philosopher-Scientist}, Library of the Living Philosophers, Evanston, IL, 1949.

\bibitem{W and Z} W.K. Wootters and W.H. Zurek, ``Complementarity in the double-slit experiment: Quantum nonseparability and a quantitative statement of Bohr's principle", \prd {\bf 19}, 473 (1979).

\bibitem{V2D2proofs} R. J. Glauber, ``Amplifiers, Attenuators, and Schr\"odinger's Cat", Ann. New York Acad. Sci. 480, 336?372 (1986); D. M. Greenberger and A. Yasin, ``Simultaneous wave and particle knowledge in a neutron interferometer", \pla{128}, 391 (1988); L. Mandel, ``Coherence and indistinguishability", \ol{16}, 1882 (1991); G. Jaeger, M. A. Horne, and A. Shimony, ``Complementarity of one-particle and two-particle interference", \pra{48}, 1023 (1993); G. Jaeger, A. Shimony, and L. Vaidman, ``Two interferometric complementarities", \pra{51}, 54 (1993); B.-G. Englert, ``Fringe Visibility and Which-Way Information: An Inequality", \prl{77}, 2154 (1996); S. D\"urr, T. Nonn, and G. Rempe, ``Fringe Visibility and Which-Way Information in an Atom Interferometer", \prl{ 81}, 5705 (1998); B.-G. Englert, M. O. Scully, and H. Walther, ``On mechanisms that enforce complementarity", J. Mod. Opt. {\bf 47}, 2213 (2000); M. Lahiri, ``Wave-particle duality and polarization properties of light in single-photon interference experiments", \pra{83}, 045803 (2011); H.-Y. Liu, J.-H. Huang, J.- R. Gao, M. S. Zubairy, and S.-Y. Zhu, ``Relation between wave-particle duality and quantum uncertainty", \pra{ 85}, 022106 (2012); F. De Zela, ``Relationship between the degree of polarization, indistinguishability, and entanglement", \pra{89}, 013845 (2014). 

\bibitem{Young} Th. Young, ``An Account of Some Cases of the Production of Colours, not Hitherto Described," Phil. Trans. Roy. Soc. London {\bf 92}, 387-397 (1802).

\bibitem{OPN}  Xiao-Feng Qian, A.N. Vamivakas and J.H. Eberly, ``Emerging Connections,  Classical and Quantum Optics", \OPN{28}, 34 (2017).

\bibitem{Spreeuw} See R.J.C. Spreeuw, ``A classical analogy of entanglement", Found. Phys. {\bf 28} 361Ð74 (1998), for a very early recognition that entanglement (non-separability of superpositions of tensor products) also belongs to classical wave theory as well as to quantum physics. 

\bibitem{B&W} The formulation we have followed is closely modeled on Chap. 10, 7th ed. of {\em Principles of Optics}, by M. Born and E. Wolf, Cambridge Univ. Press, Cambridge  (1999), especially Sec. 10.3.

\bibitem{Jaeger-etal}  G. Jaeger, M.A. Horne and A. Shimony, ``Complementarity of One-particle and Two-particle Interference" \pra{48}, 1023 (1993);  See also G. Jaeger, A. Shimony and L. Vaidman, ``Two interferometric complementarities" \pra{51}, 54 (1995).

\bibitem{Knight} P.L. Knight, ``Where the weirdness comes from", \nat{395}, 12 (1998).

\bibitem{QLHE} X.F. Qian, Bethany Little, J.C. Howell and J.H. Eberly, ``Shifting the Quantum-Classical Boundary: theory and experiment for statistically classical optical fields", \opt{2}, 611 (2015).

\bibitem{CCC} J.H. Eberly, ``Correlation, Coherence and Context", \lphys{26}, 084004 (2016).

\bibitem{Wootters} W. K. Wootters, ``Entanglement of Formation of an Arbitrary State of Two Qubits" \prl{80}, 2245 (1998). Concurrence is designed for two-qubit entanglement, and so is particularly appropriate in qubit-analog contexts such as two-slit  interference configurations. It quantifies entanglement (inseparability) and is strictly bounded, $0 \le C \le 1$.

\bibitem{Qian-etal} See X.-F. Qian, S.K. Manikandan, A.N. Vamivakas and J.H. Eberly, to be reported elsewhere.

\bibitem{James-etal} D.F.V. James, P.G. Kwiat, W.J. Munro, and A.G. White, ``On the Measurement of Qubits" \pra{64}, 052312  (2001).

\bibitem{Abouraddy-etal}  A.F. Abouraddy, K.H. Kagalwala, and B. E. A. Saleh, ``Two-point optical coherency matrix tomography" \ol{39}, 2411 (2014).

\bibitem{Jakob-Bergou} M. Jakob and J.A. Bergou, ``Quantitative complementarity relations in bipartite systems: Entanglement as a physical reality", \oc{283}, 827 (2010).




\end{thebibliography}
\end{document}